# Routes to heavy-fermion superconductivity

F Steglich[1], O Stockert[1], S Wirth[1], C Geibel[1], H Q Yuan[2], S Kirchner[1,3] and Q Si[4]

[1]Max Planck Institute for Chemical Physics of Solids, Nöthnitzer Str. 40, 01187 Dresden, Germany

[2]Department of Physics and Center for Correlated Matter, Zhejiang University, Hangzhou, Zhejiang 310027, China

[3]Max Planck Institute for Physics of Complex Systems, Nöthnitzer Str. 38, 01187 Dresden, Germany

[4]Department of Physics and Astronomy, Rice University, Houston, TX 77005, USA

steglich@cpfs.mpg.de

**Abstract**. Superconductivity in lanthanide- and actinide-based heavy-fermion (HF) metals can have different microscopic origins. Among others, Cooper pair formation based on fluctuations of the valence, of the quadrupole moment or of the spin of the localized $4f/5f$ shell have been proposed. Spin-fluctuation mediated superconductivity in $CeCu_2Si_2$ was demonstrated by inelastic neutron scattering to exist in the vicinity of a spin-density-wave (SDW) quantum critical point (QCP). The isostructural HF compound $YbRh_2Si_2$ which is prototypical for a Kondo-breakdown QCP has so far not shown any sign of superconductivity down to $T \approx 10$ mK. In contrast, results of de-Haas-van-Alphen experiments by Shishido et al. (J. Phys. Soc. Jpn. 74, 1103 (2005)) suggest superconductivity in $CeRhIn_5$ close to an antiferromagnetic QCP beyond the SDW type, at which the Kondo effect breaks down. For the related compound $CeCoIn_5$ however, a field-induced QCP of SDW type is extrapolated to exist inside the superconducting phase.

## 1. Heavy-fermion metals

Heavy-fermion (HF) metals are intermetallics containing certain lanthanide (e.g., Ce and Yb) or actinide (e.g., U and Pu) elements. The lanthanide-based HF metals are commonly regarded as prototypes of Kondo-lattice systems, for which the Doniach ($T – J$) phase diagram (figure 1) holds [2]. Here, $T$ is the absolute temperature and $J$ describes the strength of the antiferromagnetic exchange interaction between the effective spin of the localized $4f$ shell and the spins of the conduction electrons. $J$ determines both of the competing fundamental interactions: the inter-site Ruderman-Kittel-Kasuya-Yoshida (RKKY) interaction which tends to stabilize the local $4f$-electron moments as well as the on-site Kondo interaction which tends to screen them. At small $J$ ("weak coupling"), the RKKY interaction dominates, and long-range magnetic [mostly: antiferromagnetic (AF)] order forms at low temperatures. At sufficiently large $J$, where the Kondo effect prevails, the latter eventually quenches the local moments. Well below the Kondo temperature $T_K$, local Kondo singlets (i.e., $4f$-electron states entangled with conduction electron states) form. In a periodic Kondo lattice they propagate, due to the Bloch theorem, and act as charge carriers with the same quantum numbers as the non-interacting conduction electrons. Because of the extremely large on–site Coulomb repulsion, these "composite fermions" exhibit a very small Fermi velocity, which is only of the order of the velocity of sound. The effective mass of these composite charge carriers, as determined from the very large Sommerfeld coefficient of the electronic specific heat $C(T)$, $\gamma = C/T$, is correspondingly large ("heavy electrons" or "heavy fermions").

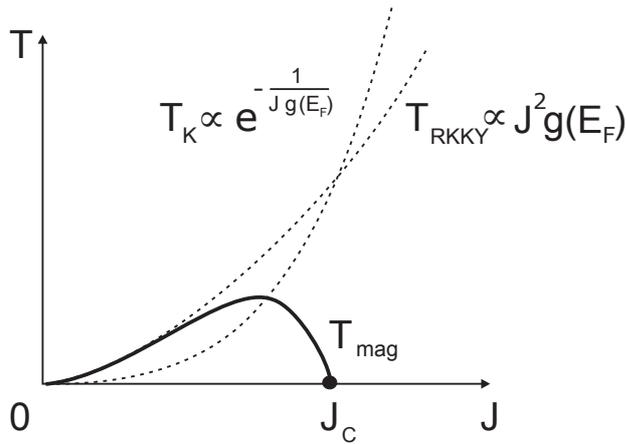

**Figure 1.** Doniach phase diagram. Binding energies for the Kondo effect, $k_B T_K$, and for the RKKY interaction, $k_B T_{RKKY}$, as a function of the 4f-conduction electron exchange integral $J$ ($J > 0$ for antiferromagnetic spin exchange). $T_{mag}$: magnetic ordering temperature ($T_{mag} \to 0$ at $J = J_c$), cf. Ref. 1.

Close to where the RKKY and Kondo interactions cancel each other ($J = J_c$, see figure 1), magnetic order can be suppressed by means of (external or chemical) pressure which influences $J$. Magnetic field, $B$, which typically causes a larger reduction of the antiferromagnetic order than that of the Kondo effect, is another non-thermal control parameter by which AF order can be suppressed. If the AF order terminates at the critical value of the control parameter in a continuous fashion, this disappearance marks a quantum critical point (QCP) [3].

Two variants of QCPs have been established for HF metals [4], depending on the behavior of the HFs upon approaching the magnetic instability at $J = J_c$ (figure 1). In both cases, these composite charge carriers are defined on the strong coupling side, where the 4f-states - being entangled with the conduction-electron states and thus constituent parts of the propagating Kondo singlets – are delocalized and contribute to a large Fermi surface.

In the first scenario, the composite charge carriers also exist on the weak-coupling side of the QCP and, therefore, AF order cannot be due to local 4f-derived magnetic moments. In this case, the magnetic order is of an itinerant nature, i.e., HF spin-density-wave (SDW) order. The QCP is then called a *SDW QCP* which denotes a continuous classical phase transition in higher dimensions [5-7]. The Fermi-surface volume at a SDW QCP stays large and undergoes a smooth change due to the gradual opening of the SDW gap. In this case, the fluctuations of the AF order parameter present the only relevant degrees of freedom. $CeCu_2Si_2$ [8] and $CeNi_2Ge_2$ [9] have been considered exemplary materials exhibiting a SDW QCP rather early.

On the other hand, if the composite fermions exist only on the paramagnetic side of the phase diagram (figure 1), the onset of AF order at the QCP is accompanied by a break-up of the Kondo singlets. On the weak-coupling side of this *Kondo-breakdown QCP* [10, 11] (strongly, incompletely yet, Kondo screened) local 4f-derived magnetic moments undergo AF order, in the presence of a small Fermi surface, made up by the conduction electrons exclusively. Consequently, the continuous AF quantum phase transition was predicted to concur with an abrupt change of the Fermi-surface volume [10, 11]. This was indeed inferred from isothermal measurements of the Hall coefficient as a function of the control parameter magnetic field, $B$, for the quantum critical material $YbRh_2Si_2$ [12, 13]. Further evidence for this unique type of QCP to be present in $YbRh_2Si_2$ is derived from the vanishing of a quantum critical energy scale $k_B T^*(B)$ (which denotes the Fermi surface crossover at finite $T$) [14] and a violation of the Wiedemann-Franz law [15]. Contrarily, $k_B T^*(B)$ remains finite at a QCP of the SDW type [4].

## 2. Variants of heavy-fermion superconductors

The advent of heavy-fermion (HF) superconductivity (SC) in $CeCu_2Si_2$ [16] followed the discoveries of superfluidity in $^3$He [17] and HF phenomena in $CeAl_3$ [18]. Given the antagonistic nature of SC and magnetism, the observation of SC in $CeCu_2Si_2$ came as big surprise: All superconductors known at that time lose their SC when doped with a tiny amount (~1 at%) of magnetic impurities – whereas in $CeCu_2Si_2$ a dense, periodic lattice of (100 at%) magnetic $Ce^{3+}$ ions is necessary to generate the superconducting state. This was inferred from the observation that the non-magnetic reference compound $LaCu_2Si_2$ is not a superconductor [16] and that SC in $CeCu_2Si_2$ is fully suppressed by doping with a tiny amount of non-magnetic impurities [19]. For $CeCu_2Si_2$, the ($T \rightarrow 0$) Sommerfeld coefficient $\gamma \approx 1$ J/K$^2$mole exceeds the $\gamma$ value of a simple metal like Cu by about three orders of magnitude. From the observation that the jump $\Delta C(T)/T$ at $T_c \approx 0.6$ K is of the same gigantic order as the normal-state value of $C/T$ at $T_c$, it was concluded [16] that the heavy-mass charge carriers make up the Cooper pairs. Similar conclusions were drawn for a few U-based HF superconductors discovered in the mid 1980ies, i.e., $UBe_{13}$ [20], $UPt_3$ [21] and $URu_2Si_2$ [22].

SC in HF metals involves pairing order parameters which are distinct from the BCS *s*-wave type. Strong support for this is lent by the existence of multiple superconducting phases, similar to what was observed for superfluid $^3$He [23]. Multiphase superconductivity was found for $UPt_3$ [24], $U_{1-x}Th_xBe_{13}$ [25] and $PrOs_4Sb_{12}$ [26]; the latter compound is unique here, since electric quadrupole fluctuations rather than magnetic dipole fluctuations, as commonly assumed for HF superconductors, are believed to mediate the Cooper pairing. Quite generally, the microscopic pairing mechanism in HF superconductors is driven by electronic interactions, contrasting SC in classical (BCS) superconductors mediated by electron-phonon coupling. In most HF superconductors, pairing is intimately related to magnetism. This was first inferred for the weak antiferromagnet $UPt_3$ ($T_N = 5$ K) from inelastic neutron-scattering (INS) spectra, which showed a drop of the magnetic intensity below $T_c = 0.5$ K [27]. For $UPd_2Al_3$, an analysis of tunnelling [28] and INS [29] results revealed that SC ($T_c \approx 2$K), which microscopically coexists with local-moment AF order ($T_N = 14.3$ K, $\mu_{ord} = 0.85$ $\mu_B$/U), is mediated by the acoustic magnon at the AF ordering wave vector [29]. Beyond these examples of SC inside AF order, SC often also occurs in the immediate vicinity of AF order. In $CeCu_2Ge_2$ [30] and $CeRh_2Si_2$ [31], SC appears near a threshold pressure $p_c$, where antiferromagnetism disappears abruptly. In many other HF metals, AF order is suppressed continuously by pressure, and SC develops near a QCP. For example, $CePd_2Si_2$ exhibits a narrow dome of SC ($T_{c,max} \approx 0.4$ K), centered around the inferred critical pressure $p_c \approx 2.8$ GPa at which AF order seems to go away smoothly [32]. Apart from an AF instability, other types of phase transitions may exist in the vicinity of SC in HF metals: $UGe_2$ [33], like $URhGe$ [34] and $UCoGe$ [35], shows SC within a regime of ferromagnetic (FM) order. Recent NMR results highlight FM fluctuations as glue for superconductivity in $UCoGe$ [36]. In $CeCu_2Si_2$ [37] and $CeCu_2Ge_2$ [38], SC is observed up to high pressure, where a weak valence instability occurs. Here, quantum critical valence fluctuations [39] are believed to mediate the formation of the Cooper pairs. In $URu_2Si_2$, SC ($T_c = 1.5$ K) coexists with some symmetry-broken state of yet unidentified origin, so-called "hidden order" ($T_{HO} = 17.5$ K) [40].

Within the last decade, the number of HF superconductors has significantly increased to about 40. The vast majority of these recently discovered HF superconductors belongs to two distinct groups of intermetallics: (i) the $Ce_nT_mIn_{3n+2m}$ family of compounds (T: transition metal), like $CeCoIn_5$ [41] and (ii) systems lacking a center of inversion symmetry, like $CePt_3Si$ [42].

The former systems are quasi-two-dimensional (2D) variants of the cubic superconductor $CeIn_3$ ($T_c \approx 0.2$ K) [32]. They are formed by stacking alternating n layers of $CeIn_3$ and m layers of $TIn_2$ sequently along the tetragonal *c*-axis. As predicted in Ref. 43, the so reduced dimensionality causes an increase of $T_c$ by more than one order of magnitude ($T_c = 2.3$ K for $CeCoIn_5$). A further substantial $T_c$-enhancement was achieved upon replacing Ce (with localized 4*f* shell) by Pu (whose 5*f* shell is spatially more extended). Among HF superconductors, $PuCoGa_5$ with $T_c = 18.5$ K is presently the record holder [44]; its Rh homologue [45] as well as $NpPd_2Al_5$ [46] also show enhanced $T_c$ values.

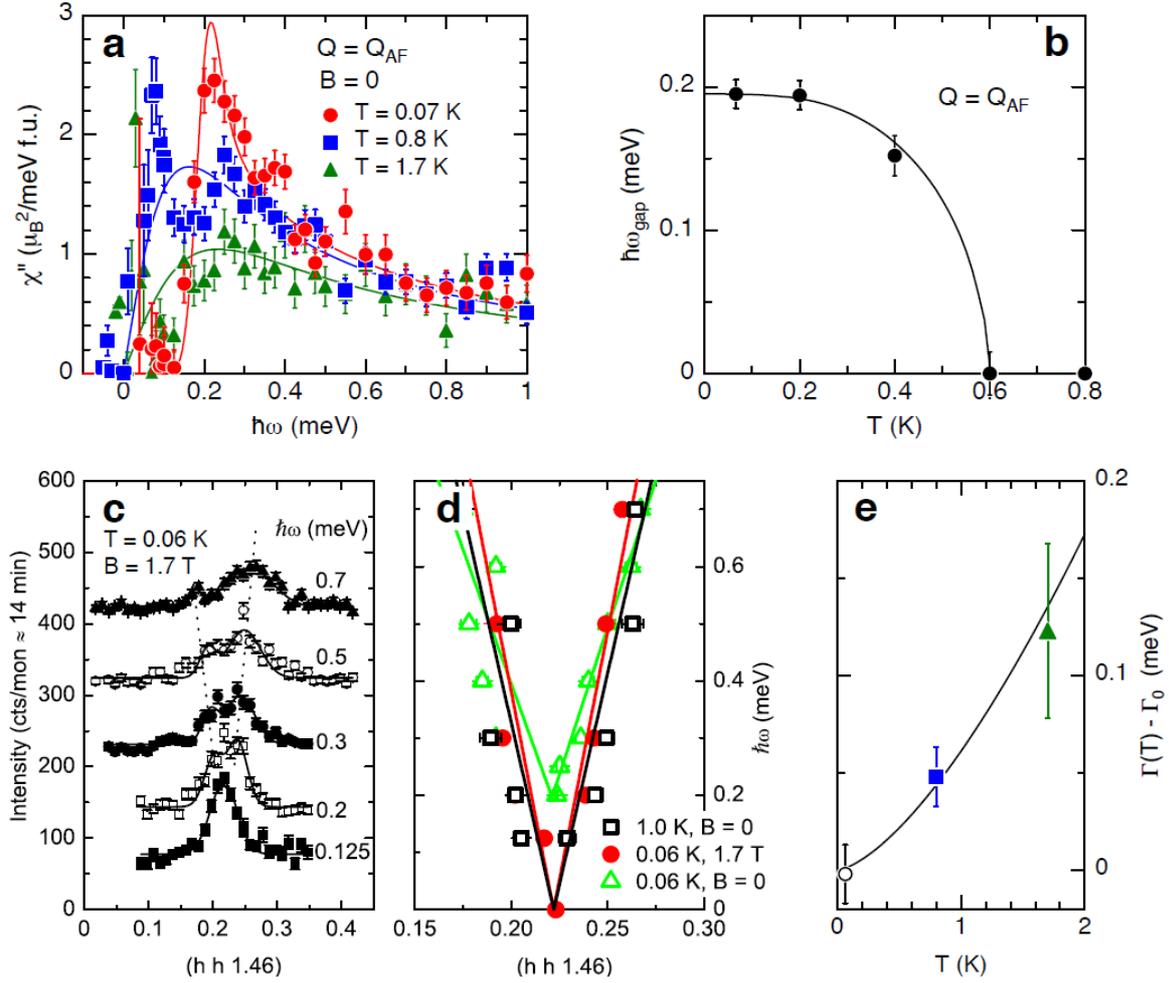

**Figure 2**. Magnetic response, relaxation rate and spin gap at the SDW ordering wave vector $Q_{AF}$ in S-type $CeCu_2Si_2$. **a** $\chi''$ vs $\hbar\omega$ for different temperatures. From Ref. 53. **b** Temperature dependence of the spin-excitation gap in the superconducting state, together with the scaled gap function for a weak-coupling d-wave superconductor (solid line). From Ref. 53. **c** Momentum dependence in the normal state for different energy transfers $\hbar\omega$ at $T = 0.06$ K, $B = 1.7$ T. Scans are shifted by 100 counts with respect to each other. Solid lines indicate fits with two Gaussian peaks; dashed lines are guides to the eye. From Ref. 54. **d** Spin fluctuation dispersion as derived from fits to $Q$ scans in the normal state at $B = 0$, $T = 1.0$ K or at $T = 0.06$ K and $B = 1.7$ T as well as in the superconducting state at $T = 0.06$ K and $B = 0$. Solid lines indicate linear fits. From Ref. 54. **e** Linewidth of the quasielastic magnetic response in the normal state, as obtained by fitting the data in **a** (lines therein). The solid line represents a $T^{1.5}$ dependence. The $T = 0$ offset is 0.112 meV. From Ref. 53.

In the past few years, the group of "non-centrosymmetric" superconductors has attracted much theoretical interest. This derives from the fact that the lack of inversion symmetry allows for a mixing of even-parity ($S = 0$) and odd-parity ($S = 1$) pair states, the degree of mixing depending on the strength of the antisymmetric spin-orbit coupling [47, 48].

The majority of HF superconductors *with* inversion symmetry show a highly anisotropic, even-parity order parameter, while a small number of them are prime candidates for odd-parity pairing [33-35, 49, 50]. Interestingly, in each of these latter cases SC coexists with either AF or FM order, cf. Refs. [27] and [33], respectively.

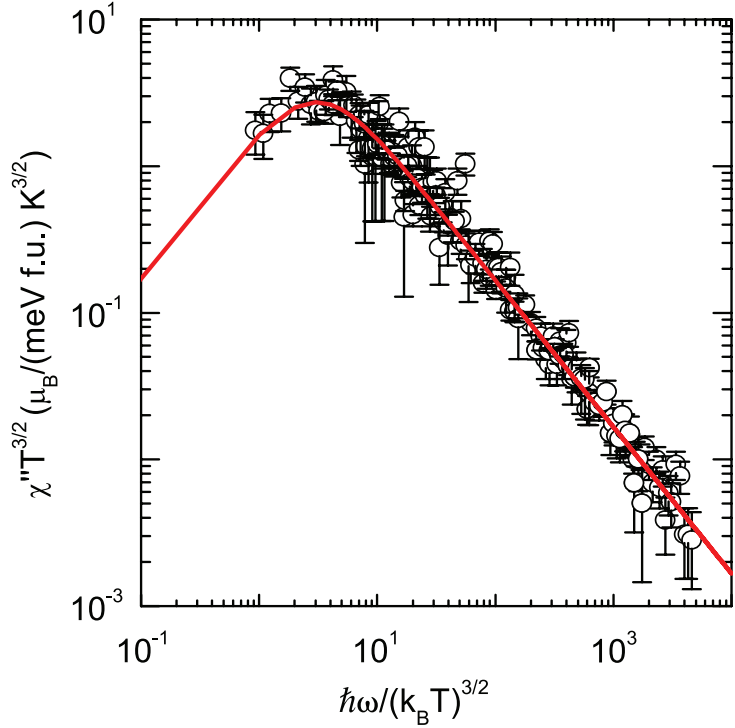

Figure 3. Universal scaling of the imaginary part of the dynamical susceptibility $\chi''$ ($\mathbf{Q}, \omega$) as a function of $\hbar\omega/(k_B T)^{3/2}$ for S-type $CeCu_2Si_2$. From Ref. 54.

## 3. Superconductivity near antiferromagnetic quantum critical points

Many of the HF superconductors exhibit a non-Fermi-liquid (nFL) low-$T$ normal state, highlighting a nearby AF QCP. For most of these nFL superconductors the QCP occurs at high pressure, $CePd_2Si_2$ being a well-known example [32]. This makes it challenging to unravel the magnetic structure as well the spectrum of critical fluctuations near the QCP by neutron scattering. $CeCu_2Si_2$ is well suited for such an investigation as here, the QCP is accessible already at ambient pressure. It is located very closely to the true stoichiometry point inside the narrow homogeneity range of the 122 phase in the chemical phase diagram [51]. Within this range, homogeneous single crystals can be grown with, e.g., a tiny deficiency or a tiny excess of Cu. The former crystals show weak AF order ("A-type"), the latter ones are superconducting ("S-type").

Neutron-diffraction measurements on A-type $CeCu_2Si_2$ revealed HF-SDW order below $T_N \approx 0.8$ K, with a tiny staggered moment of $\approx 0.1$ $\mu_B$/Ce and an incommensurate ordering wave vector $\mathbf{Q}_{AF}$ = (0.215 0.215 0.53) [52].

Recently, INS experiments have been performed on an S-type $CeCu_2Si_2$ single crystal [53]. As shown in figure 2a, this sample exhibits a broad quasielastic magnetic response in its low-temperature normal state. SC induces a gap below 0.2 meV in the quasielastic spectrum. This gap cannot be related to the opening of a SDW gap, as it is absent in A-type $CeCu_2Si_2$. It rather refers to the opening of a superconducting gap in the HF density of states at the Fermi level (figure 2b). The quasielastic response is extremely localized in $\mathbf{Q}$-space, i.e., only around the incommensurate ordering wave vector $\mathbf{Q}_{AF}$ where SDW order occurs nearby in the phase diagram (figures 2c,d). This clearly shows that the spectra in figure 2a represent a broad distribution of dynamical SDW correlations. Their lifetime strongly increases upon lowering the temperature ("slowing down"), cf. figure 2e, indicating the close proximity of AF order. As shown in figure 3, these "almost quantum critical spin fluctuations" obey a universal scaling over wide ranges of energy transfer and temperature, as expected for 3D critical modes associated with a SDW QCP [54].

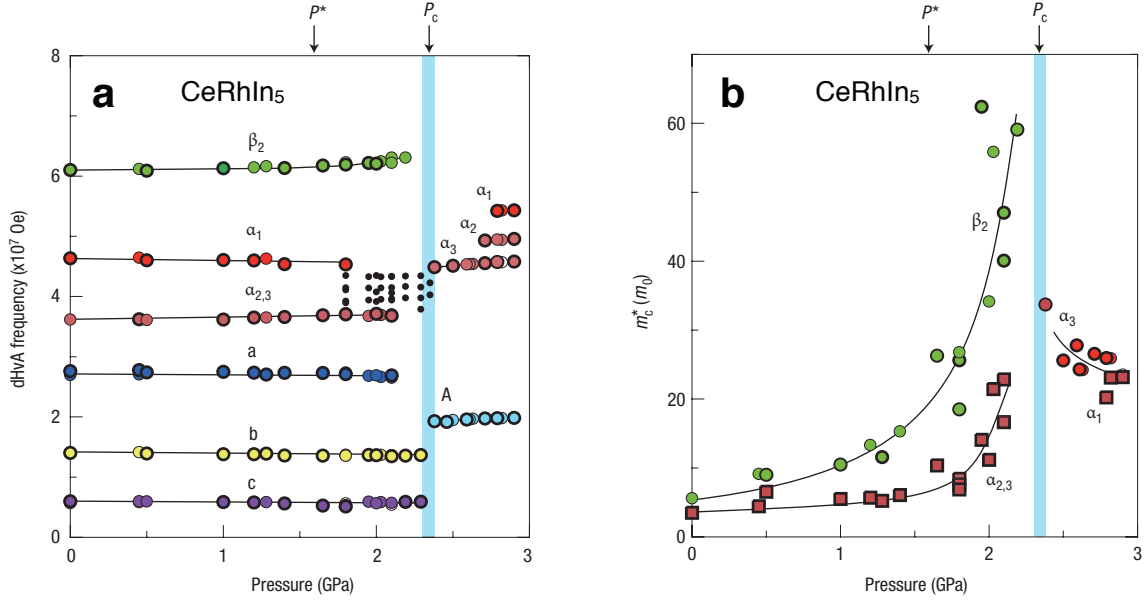

Figure 4. Changes of Fermi-surface properties across a pressure-induced Kondo-breakdown quantum critical point in CeRhIn$_5$. Pressure dependence of de Haas-van Alphen frequencies (**a**) and cyclotron masses (**b**). From Ref. 61.

In $Q$-space, the magnetic response of S-type CeCu$_2$Si$_2$ discussed before appears as an overdamped, dispersive mode ("AF paramagnon") [53]. From the slope of the dispersion relation in the normal state (figure 2d), the paramagnon velocity $v_{pm}$ was found to be smaller by almost one order of magnitude than the Fermi velocity $v_F^*$ of the composite charge carriers [53]. As mentioned in Sect. 1, $v_F^*$ is of the order of the velocity of sound only [55], which prevents retardation of the electron-phonon coupling. On the other hand, because $v_F^* \gg v_{pm}$, the coupling between the heavy fermions and the SDW fluctuations is well retarded. It implies that in this interaction the direct Coulomb repulsion between the charge carriers is avoided and the magnetic excitations can provide the glue for SC. This is also concluded from the large difference between the spectral weight in the normal and superconducting states, which highlights a huge saving in magnetic exchange energy, when compared to the superconducting condensation energy [53]. The latter is reliably determined from the specific heat both in the superconducting and field-driven normal state. The exchange energy saving exceeds the condensation energy by a factor larger than 20. Correspondingly, the loss of kinetic energy exceeds the condensation energy by also a factor of order 20 – compared to a factor of only 2- 3 in the case of classical (BCS) superconductors. This observation is understood if there is a break-up of the Kondo effect at energies above a relatively small Kondo-breakdown (Fermi-surface crossover) energy scale $k_B T^*$, which causes a shift of spectral weight in the electron spectrum from below $k_B T^*$ to above it and up to several eV, i.e., the order of the local Coulomb interaction. I.o.w., even for this canonical example of SC near an SDW QCP, the f-electron localization comes into play dynamically.

For a HF metal like CeCu$_2$Si$_2$, this Fermi-surface crossover temperature $T^*$, which generically is below $T_K$ = 15-20 K, must be finite [56]. With respect to the magnetic response in the INS spectra shown in figure 2a, which extends to about 2 meV, corresponding to $k_B T_K$, this implies that only the low-frequency spinfluctuations involved in the Cooper pairing are of the collective paramagnon type, while those with frequencies larger than $k_B T^*$ must be related to the Kondo effect.

In the future, searching for a Kondo-breakdown transition at which $T^* \rightarrow 0$ (at a critical value of the control parameter) inside the magnetically ordered phase of HF antiferromagnets will be of high timely interest. Measurements under both chemical [57] and hydrostatic [58] pressure have suggested the existence of such a Kondo-breakdown transition in volume compressed YbRh$_2$Si$_2$. Very recent Fermi-surface studies via measurements of the de-Haas-van Alphen effect under high magnetic fields have identified a Kondo-breakdown transition inside the AF phase in CeRhIn$_5$: It occurs near $B^* \approx 40$ T, i.e., well below the critical field, where AF order is smoothly suppressed at absolute zero temperature, $B_N \approx 50$ T [59]. For magnetic fields $B < B^*$ and at $p = 0$, CeRhIn$_5$ is, therefore, a local-moment HF antiferromagnet. This holds true also at lower fields and finite pressure. In the field range 10 – 17 T and near the critical pressure $p_N = 2.3$ GPa where $T_N \to 0$ [60], de Haas-van Alphen oscillations indicate an abrupt reconstruction of the Fermi surface (figure 4a) that is accompanied by an incipient divergence of the cyclotron mass (figure 4b) [61]. Therefore, also at finite pressure, AF order at $B = 0$ is very likely of the local-moment type. Further on, the putative AF QCP masked by the pressure-induced superconducting dome in CeRhIn$_5$ is most probably of the Kondo-breakdown variety. This suggests that HF SC not only arises in the vicinity of SDW QCPs, like in CeCu$_2$Si$_2$ [53, 54], but can also be driven by the purely electronic fluctuations of the Kondo-breakdown QCP.

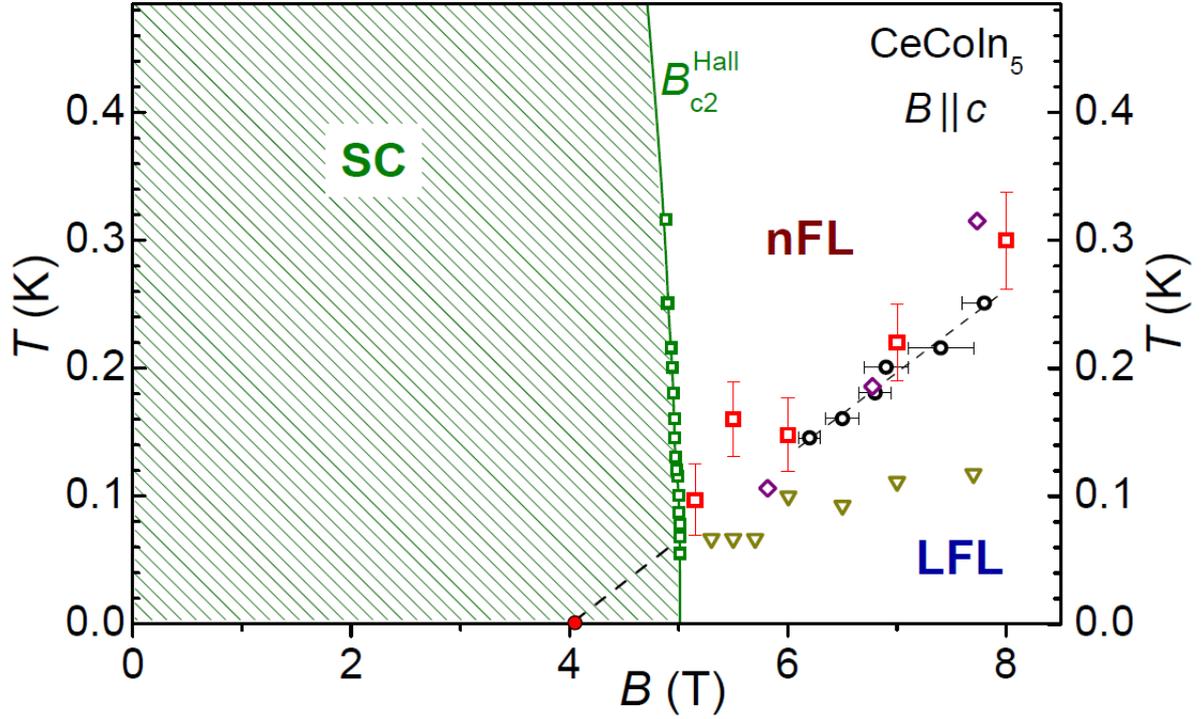

**Figure 5**. Temperature-magnetic field ($T$-$B$) phase diagram of CeCoIn$_5$ for $B \parallel [001]$. The black circles mark results from Hall measurements [64] indicating the cross-over from Landau Fermi-liquid (LFL) to non-Fermi-liquid (nFL) behavior. An extrapolation hints at a putative QCP at $B_{QCP} \sim 4.1$ T(red dot), i.e,. inside the superconducting phase (SC). Results from thermal expansion (red squares, [65]) and early magnetoresistance (purple diamonds, [66]) measurements point toward a similar value. the deviation from $T^2$-behavior in the longitudinal magnetoresistance (triangles, [67]) was also interpreted as $B_{QCP} < B_{c2,0}$.

It is interesting to compare CeRhIn$_5$ with its Co homologue, when the field is applied along the c-axis. Early reports on the magneto-resistance [62] and specific heat [63] of CeCoIn$_5$ claimed the AF QCP to coincide with the upper critical field at $T = 0$, i.e., $B_{QCP} \approx B_{c2,0}$, while later reports point toward a putative QCP inside the superconducting phase [64-66], with $B_{QCP} \approx 4.1$ T (see figure 5). In subsequent current-voltage measurements inside the Shubnikov phase of superconducting CeCoIn$_5$, Hu et al. observed a sharp increase in the flux-flow resistivity upon decreasing either the temperature or the magnetic field [68]. This increase was ascribed to quasiparticle scattering off of critical AF fluctuations and, consequently, provides a strong indication for the existence of an AF phase boundary in the *B-T* phase diagram, i.e., below $T = 1.6$ K and $B = B_{QCP} = 4.1$ T, in excellent agreement with the results of [64, 65, 67]. Also, a transition very recently observed at roughly 4 T in the isothermal field-dependent entropy was suggested to be related to a QCP hidden by the superconducting phase [69]. The latter findings [69] as well as thermal expansion studies [70] are compatible with SDW order – in contrast to CeRhIn$_5$ where AF order is of the local-moment type, except for magnetic fields $B > B^* \approx$ 40 T [59].

**4. Perspective**
HF superconductors provide a multitude of unconventional SC scenarios. Most frequently, SC in close vicinity of AF QCPs has been established. As proposed in, e.g., Ref. 71 and 43, HF SC near a SDW QCP, mediated by almost quantum critical SDW fluctuations, could indeed be verified for CeCu$_2$Si$_2$ [53, 54]. It will be important to experimentally check whether this SDW QCP scenario can be applied to other (pressure-induced) nFL superconductors, such as CePd$_2$Si$_2$ [32]. Pressure-induced SC in CeRhIn$_5$ [72] appears to occur [60, 61] near a Kondo-breakdown QCP, which is often labeled a zero-temperature 4*f*-selective Mott transition. This offers a link between HF SC and unconventional SC in other families of strongly correlated electron systems, including the newly discovered Fe-based pnictides/chalcogenides and the doped Mott insulators of the cuprates and organic charge transfer salts, cf. various contributions to this conference.

**Acknowledgements**
Part of the work done in Dresden was supported by the DFG Research Unit 960, "Quantum Phase Transitions". HQY is partially supported by NSFC, the National Basic Research Program of China (973 Program) and the Zhejiang Provincial Natural Science Foundation of China. The Work at Rice University has in part been supported by NSF Grant No. DMR-1006985 and the Robert A. Welch Foundation Grant No. C-1411.